\documentstyle[12pt,epsf]{article}
\newcommand{\nc}{\newcommand}
\def\frac#1#2{{\textstyle {#1 \over #2}}}

\nc{\beq}{\begin{equation}}
\nc{\eeq}{\end{equation}}
\nc{\beqa}{\begin{eqnarray}}
\nc{\eeqa}{\end{eqnarray}}
\nc{\lsim}{\begin{array}{c}\,\sim\vspace{-21pt}\\< \end{array}}
\nc{\gsim}{\begin{array}{c}\sim\vspace{-21pt}\\> \end{array}}
\def\NN{\hbox{\it I\hskip -2.pt N}}
\def\ZZ{\hbox{\it Z\hskip -4.pt Z}}

\def\D{{\cal D}}
\def\noi{\noindent}

\def\diagram#1{\def\normalbaselines{\baselineskip=0pt
\lineskip=10pt\lineskiplimit=1pt} \matrix{#1}}

\def\hfl#1#2{\smash{\mathop{\hbox to 12mm{\rightarrowfill}}
\limits^{\scriptstyle#1}_{\scriptstyle#2}}}

\def\hfr#1#2{\smash{\mathop{\hbox to 12mm{\leftarrowfill}}
\limits^{\scriptstyle#1}_{\scriptstyle#2}}}

\def\RR{\hbox{\it I\hskip -2.pt R }}
\def\CC{\hbox{\it l\hskip -5.5pt C\/}}

\def\cc#1{\kern .7em\hfill #1 \hfill\kern .7em}

\newcommand{\mysection}[1]{\setcounter{equation}{0}\section{#1}}

\begin{document}
\begin{titlepage}

\begin{center}
July, 2000      \hfill       PM/00-20\\
\vskip .3 in {\large \bf Fractional Supersymmetry~ and}\\
{\large \bf Lie Algebras}\\
\vskip .3truecm
{
  {\bf M. Rausch de Traubenberg}\footnote{rausch@lpm.univ-montp2.fr,
rausch@lpt1.u-strasbg.fr}
   \vskip 0.2 cm
   {\it  Laboratoire de Physique Th\'eorique, Universit\'e Louis Pasteur}\\
   {\it 3-5 rue de l'universit\'e, 67084 Strasbourg Cedex, France}\\ 
  \vskip 0.2 cm
 and \\
   {\it Laboratoire de Physique Math\'ematique et Th\'eorique,
         Universit\'e de Montpellier 2}\\
   {\it place Eug\`ene Bataillon, case 70, 34095 Montpellier Cedex 5, 
     France}\\ 
}

\end{center}

\vskip .5 in
\begin{abstract}
Supersymmetry and super-Lie algebras have been consistently generalized
previously.
The so-called fractional supersymmetry and $F-$Lie algebras could be 
constructed starting from any representation $\D$ of any Lie algebra $g$.
This involves taking the $F^{\mathrm th}$ root of $\D$ in some sense.
We show,  after having constructed differential realization(s) of  any
Lie algebra, how  fractional supersymmetry 
can be explicitly realized in terms of
appropriate homogeneous monomials.
\end{abstract}

\vskip .7 truecm

\noi
{\it
\begin{center}
Lecture given at the Workshop on Non Commutative Geometry and
Superstring Theory,

 Rabat 16-17 June 2000.
\end{center}
} 
\end{titlepage}

\mysection{Introduction}
Describing   the laws of physics in terms of  underlying symmetries has 
always been a powerful tool. In this respect, it is interesting to study
the kind of symmetries which are allowed in space-time. Within the
framework of Quantum Field Theory (unitarity of the $S$ matrix {\it etc}) 
it is generally admitted that we cannot go beyond supersymmetry (SUSY).
However,
the no-go theorem stating that supersymmetry is {\it the only non-trivial
extension beyond the  Poincar\'e algebra} is valid only if one considers
Lie or Super-Lie algebras. Indeed, if one considers Lie algebras, 
the Coleman and Mandula theorem \cite{cm}  allows only trivial extensions
of the Poincar\'e symmetry, {\it i.e.} extra symmetries must
commute with the Poincar\'e 
generators.
In contrast, if we consider superalgebras, the theorem of 
Haag, Lopuszanski and Sohnius \cite{hls} shows that we can construct a
unique (up to the number of supercharges) superalgebra
extending the Poincar\'e Lie algebra  non-trivially.  It may seem that
these two theorems encompass all possible symmetries of space-time.
But, if one examines the hypotheses of the above theorems, one sees that it
is possible to imagine symmetries which go beyond supersymmetry. Several
possibilities have been considered in the literature \cite{ker, luis,
fsusy, fsusy1d, fr,am, prs, fsusy2d, fvir, fsusy3d, fsusyh}, 
the intuitive idea being
that the generators of the Poincar\'e algebra are obtained as an appropriate
product of more fundamental additional symmetries. These new generators
are in a representation of the Lorentz group which can be neither
bosonic nor fermionic (bosonic charges close under commutators and 
generate a Lie algebra, whilst fermionic charges close under anticommutators
and  induce  super-Lie algebras). 
In an earlier work we proposed an algebraic structure,
called an $F-$Lie algebra, which makes this idea
precise in the context of fractional supersymmetry (FSUSY) of order $F$
\cite{vg}.
Of course, when $F=1$ this is a Lie algebra, and when 
$F=2$ this is  a Super-Lie algebra.
Within the framework of this algebraic structure, we showed that starting from
any representation ${\cal D}$ of any Lie algebra $g$ it is possible to 
take in some sense the $F^{\mathrm{th}}-$root of ${\cal D}$. This means that
we were able to consider a representation ${\cal D^\prime}$ such that
the symmetric tensorial product of order $F~~$ 
${\cal S}^F\left({\cal D^\prime}\right)$
is related to ${\cal D}$ \cite{vg}. The representation ${\cal D^\prime}$ is in
general an infinite dimensional representation of $g$, {\it i.e.}
a Verma module \cite{kr}. The purpose of
this note is to give an explicit way to realize this construction in
terms of monomials, 
{\it i.e.} using appropriate differential realizations of $g$. \\

The content of this paper is as follow. In section 2, we summarize the
results of \cite{vg} about $F-$Lie algebras. In section 3 we show how
for any Lie algebra we can realize the representations in terms of 
homogeneous monomials constructing differential realization(s) of
the algebra $g$. Finally, in section 4 FSUSY is realized
for the algebra $g=su(3)$. Of course this construction works along the 
same lines for any Lie algebra.

\mysection{Algebraic Structure of Fractional Supersymmetry}

In this section, we recall the abstract mathematical structure 
which generalizes the theory of Lie super-algebras
and their (unitary) representations. Let $F$ be a positive integer and 
$q=\exp{({2i \pi \over F})}$.
We consider a complex vector space $S$ together with a linear map $\varepsilon$
from $S$ into itself satisfying $\varepsilon^F=1$. 
We set $A_k= S_{q^k}$ and $B=S_1$  (where $S_\lambda$  is the eigenspace 
corresponding to the eigenvalue $\lambda$ of $\varepsilon$)
so that $S=B\oplus_{k=1}^{F-1} A_k$. The map $\varepsilon$ is called the 
grading.
If $S$ is endowed with the following structures we will say that $S$ is
a fractional super Lie algebra ($F$-Lie algebra for short):

\begin{enumerate}
\item $B$ is a Lie algebra and $A_k$ is a   representation of $B$.
\item There are multilinear,  $B-$equivariant  ({\it i.e.} which respect 
the action  of  $B$) maps
 $\left\{~~, \cdots,~~ \right\}: {\cal S}^F\left(A_k\right) 
\rightarrow B$ from 
${\cal S}^F\left(A_k\right)$ into  $B$.
In other words, we  assume that some of the elements of the Lie algebra $B$ can
be expressed as $F-$th order symmetric products of
``more fundamental generators''. Here  ${ \cal S}^F(D)$ denotes 
the $F-$fold symmetric product
of $D$. It is then easy to see that: 

\beqa
\label{eq:epsi}
\left\{\varepsilon(a_1), \cdots, \varepsilon(a_F)\right\}=
\varepsilon\left(\left\{a_1, \cdots, a_F\right\}\right), 
\forall a_i \in A_k.
\eeqa

\item  For $b_i \in B$ and $a_j \in A_k$ the following 
``Jacobi identities'' hold:

\beqa
\label{eq:J}
&&\left[\left[b_1,b_2\right],b_3\right] + 
\left[\left[b_2,b_3\right],b_1\right] +
\left[\left[b_3,b_1\right],b_2\right] =0 \nonumber \\
&&\left[\left[b_1,b_2\right],a_3\right] +
\left[\left[b_2,a_3\right],b_1\right] +
\left[\left[a_3,b_1\right],b_2\right]  =0 \nonumber \\
&&\left[b,\left\{a_1,\dots,a_F\right\}\right] =
\left\{\left[b,a_1 \right],\dots,a_F\right\}  +
\dots +
\left\{a_1,\dots,\left[b,a_F\right] \right\} \\
&&\sum\limits_{i=1}^{F+1} \left[ a_i,\left\{a_1,\dots,
a_{i-1},
a_{i+1},\dots,a_{F+1}\right\} \right] =0. \nonumber
\eeqa

\noindent
The first identity is the usual Jacobi identity for Lie algebras,
the second says that the $A_k$ are representation spaces of $B$ and
the third is just the Leibniz rule (or the equivariance of  
 $\left\{~~, \cdots,~~ \right\}$). The fourth identity is
the analogue of the graded Leibniz rule of Super-Lie algebras
for $F-$Lie algebras

\hskip -1 cm 
If we want to be able to talk about unitarity, we also require the
following additional 
struc-

\hskip -1 cm
ture and in this case, $S$ is called an
$F-$Lie algebra with adjoint.

\vfill \eject
\item
A conjugate linear map $\dag$ from $S$ into itself  such that:

\beqa
\label{eq:conj}
\begin{array}{ll}
\mathrm{a)}&(s^\dag)^\dag=s, \forall s \in S \cr
\mathrm{b)}& \left[a,b\right]^\dag= \left[b^\dag,a^\dag\right] \cr 
\mathrm{c)}&\varepsilon(s^\dag)=\varepsilon(s)^\dag \cr
\mathrm{d)}& \left\{a_1,\cdots,a_F\right\}^\dag=
 \left\{\left(a_1\right)^\dag,\cdots,\left(a_F\right)^\dag\right\},~~~
\forall a \in A_k.
\end{array}
\eeqa

From a) and c) we see that for 
$X \in B$ we have  $X^\dag \in B$, and that for
  $X \in A_k$, we have   $ X^\dag \in  A_{F-k}.$
\end{enumerate} 
\noindent

A unitary representation of an $F-$Lie algebra
with adjoint $S$  is  a linear map
$\rho : ~ S \to \mathrm{End}(H)$, 
(where $H$ is a Hilbert space
and ${\mathrm{End}}(H)$ the space of linear operators acting on $H$) 
and a unitary
endomorphism $\hat \varepsilon$ such that $ \hat \varepsilon^F=1$ 
which satisfy 

\beqa      
\label{eq:rep}
\begin{array}{ll}
\mathrm{a)}& \rho\left(\left[x,y\right]\right)= \rho(x) \rho(y)- 
\rho(y)\rho(x) \cr
\mathrm{b)}& \rho \left\{a_1.\cdots,a_F\right\}=
{1 \over F !} \sum \limits_{\sigma \in S_F}
\rho\left(a_{\sigma(1)}\right) \cdots \rho\left(a_{\sigma(F)}\right) \cr
\mathrm{c)}& \rho(s)^\dag = \rho(s^\dag) \cr
\mathrm{d)}& \hat \varepsilon \rho\left(s\right) \hat \varepsilon^{-1} =
\rho\left(\varepsilon\left(s\right)\right)
\end{array}
\eeqa

\noindent
($S_F$ being the group of permutations of $F$ elements).
As  a consequence of these properties, 
since the eigenvalues of $\hat \varepsilon$ are $\mathrm{F}^{\mathrm{th}}-$
roots of unity, we have  the following decomposition of the Hilbert space

$$H= \bigoplus \limits_{k=0}^{F-1} H_k,$$

\noindent
where $H_k=\left\{\left|h\right> \in H ~:~ 
\hat \varepsilon\left|h\right>=q^k \left|h\right> \right\}$.
The operator $N \in \mathrm{End}(H)$
defined by $N\left|h\right>=k
\left| h \right>$ if $\left|h\right> \in H_k$
is the  ``number operator'' (obviously $q^N=\hat \varepsilon$).
Since $\hat \varepsilon \rho(b)= \rho(b) \hat \varepsilon, \forall b \in B$
each $H_k$ provides a representation of the Lie algebra $B$. 
Furthermore, for $a \in A_\ell$,
 $\hat \varepsilon \rho(a)=q^\ell \rho(a) \hat \varepsilon$ and so
we have 
$\rho(a) .H_k\ \subseteq 
H_{k+\ell ({\mathrm{mod~} F)}}$  \\

Several remarks can be made at that point
Firstly, for all $k=1,\cdots, F-1$ it is clear that 
the subspace  $B \oplus A_k$ of $S$ satisfies (\ref{eq:epsi}-\ref{eq:J}) 
and the subspace $B \oplus A_k \oplus A_{-k}$ satisfies  
(\ref{eq:epsi}-\ref{eq:conj}) (when $S$ has an adjoint). 
Secondly, it is important to notice
that bracket $\{ \cdots \}$ is a priori  not defined for  elements in different
gradings. \\

The basic idea to define a fractional supersymmetry is the following.
Let $g$ be a  Lie algebra and let ${\cal D}, {\cal D^\prime}$
be  representations of $g$. The representation ${\cal D^\prime}$
is chosen in such a way that ${\cal S}^F \left({\cal D}^\prime\right)$
is related to ${\cal D}$ 
(${\cal S}^F \left({\cal D}^\prime\right) \sim {\cal D}$)
in a sense specified later on.
Then we consider $B=g \oplus {\cal D}$, a  Lie algebra 
as the semi-direct product of $g$ 
and $A_1= {\cal D}^\prime$.  
The relation ${\cal S}^F \left({\cal D}^\prime\right) \sim {\cal D}$
will be the fundamental one to define an $F-$Lie algebra.

\mysection{Differential realization(s) of Lie algebras}

We consider now $g$ a  semi-simple Lie algebra 
of rank $r$.
Let
$h$ be a Cartan sub-algebra of $g$, let $\Phi \subset h^\star$ (the dual of
$h$)  be the corresponding set of roots and let
$f_\alpha$ be the one dimensional root space associated to $\alpha \in \Phi$.
We chose a basis  $\{H_i, i=1, \cdots,  r\}$ of $h$ and elements 
$E^\alpha \in f_\alpha$ such that the commutation relations  become

\beqa
\label{eq:lie}
\big[H_i,H_j \big] &=& 0 \nonumber \\
\big[H_i,E^\alpha \big] &=&  \alpha^i E^\alpha \\
\big[E^\alpha, E^\beta\big] &=& \left \{
\begin{array}{ll}
\epsilon\{\alpha,\beta\} E^{\alpha+\beta}& {\mathrm {~~if~~}} 
\alpha + \beta \in \Phi \cr
{2\alpha.H \over \alpha.\alpha}& {\mathrm {~~if~~} } \alpha+\beta=0 \cr
0& {\mathrm {~~otherwise }}
\end{array}
\right. \nonumber
\eeqa

We now introduce $\{\alpha_{(1)},\cdots, \alpha_{(r)}\}$ 
a basis of simple roots.  
The weight lattice $\Lambda_W(g) \subset h^\star$ is the set of vectors
$\mu$ such that $ {2 \alpha.\mu \over \alpha.\alpha} \in \ZZ$ and, as is
well known, there is a  basis of the weight lattice consisting of the
fundamental weights $\{\mu_{(1)}, \cdots, \mu_{(r)} \}$ defined by
$ {2 \mu_{(i)}.\alpha_{(j)} \over \alpha_{(j)}.\alpha_{(j)}}= \delta_{ij}$.
A weight $\mu = \sum \limits_{i=1}^r n_i \mu_{(i)}$ is called dominant if
all the $n_i \ge 0$ and it is well known that the set of dominant weights
is in one to one correspondence with the set of (equivalence classes of)
irreducible finite dimensional representations of $g$.

Recall briefly how one can associate a  representation of $g$ to 
$\mu \in h^\star$. A given representation of $g$ can be defined from
a highest weight states $| \mu >$ ($E^\alpha | \mu > =0, \alpha >0, 
2\frac{\alpha_{(i)}.H}{\alpha^2} | \mu> = n_i | \mu>), i=1,\cdots,r $).
We denote $h_i=2\frac{\alpha_{(i)}.H}{\alpha^2}$.
The space obtained from $| \mu >$ by  the action of the
element of $g$:~ 
$E^{-\alpha_{(i_1)}} \cdots E^{-\alpha_{(i_k)}} | \mu>$
clearly define a representation of $g$. This construction
can be made more precise using the language of Verma module \cite{vg, kr}.
This representation is denoted $\D_\mu$.
To come back to our original  problem, consider a finite dimensional 
irreducible representation $\D_\mu$. The basic idea to define an FSUSY
associated to $g \oplus \D_\mu$ is to consider the infinite dimensional
representation associated to the weight $\mu/F$. In \cite{vg},  
we defined an $F-$Lie
algebra associated to $g, \D_\mu$ and $\D_{\mu/F}$. 
Here, we reproduce these results in an explicit way
using the differential realization of $g$.\\

As we have recalled the representations of $g$ are just specified by
the weight $\mu = \sum \limits_{i=1}^r n_i \mu_{(i)}$. But, 
among the representations  of $g$ there are $r$ basic representations. 
These representations are associated to the fundamental weights
 $\mu=\mu_{(i)}$,
and all representations can be obtained from (symmetric) tensorial product of 
these basic representations. Furthermore, all basic representations can be 
derived from the antisymmetric product of the elementary representations,
which are associated to the weight with terminal point in the Dynkin
diagram \cite{w}. Consequently, if one is able to obtain a differential
realization  of the elementary representations of $g$ (at maximum
$3$), one is able to construct { \it all representations} easily as we 
will see. Moreover, for $su(n+1)=a_n, sp(2n)=c_n$, we need to
consider  only the fundamental representation (related to $\mu=\mu_{(1)}$).
Although for $so(2n+1)=b_n$ the vector ($\mu=\mu_{(1)}$) and the spinorial
($\mu=\mu_{(n)}$) representations and for $so(2n)=d_n$ the vector 
($\mu=\mu_{(1)}$) and the two spinorial representations ($\mu=\mu_{(n-1)}, 
\mu=\mu_{(n)}$)  reproduce all representations. For the exceptional Lie
algebras some simplifications may happen  from the embedding properties
($e_8 \subset so(16), e_7 \subset so(12) \oplus su(2), e_6 \subset so(10)
\oplus u(1)$), and $f_4, g_2$ being of small rank calculations can be done
easily.
In the next subsection we just consider the series
$a_n,b_n,c_n$ and $d_n$ and construct finite and infinite dimensional
representations. 
It is important to emphasize that all the highest weight representations,
finite and infinite dimensional, can be obtained in terms of homogeneous
monomial of appropriate variables as a consequence of the differential 
realization of $g$.
In the next section we apply these 
realizations to construct explicitly FSUSY.

\subsection{$su(n)$}
$su(n)$ is a rank $n-1$ Lie algebra, but it is more convenient for our
purpose to define roots as vectors of $\RR^n$. Introduce $e_i, i=1,\cdots,n$
an orthonormal basis of $\RR^n$ the simples roots of $su(n)$ reads

\beq
\label{eq:simple_sun}
\alpha_{(i)}=e_i-e_{i+1}, \ \ i=1,\cdots, n-1,
\eeq

\noi
and the positive roots  

\beq
\label{eq:roots_sun}
\alpha=e_i-e_j, \ \ 1 \le i < j \le n.
\eeq

Now if we introduce\footnote{
From now on, all  the variables, whatever $su(n), so(n)$ or $sp(2n)$
are concerned, are positive and different from zero.}
$(x_1, \cdots, x_n) \in (\RR^+-\{0\})^n$ we can define
the explicit realization of $su(n)$: 

\beqa
\label{eq:sun}
E^\alpha &=& E^{e_i-e_j} = x_i \partial_{x_j},   \ \ 1 \le i < j \le n \\
h_i&=& x_i \partial_{x_i}-x_{i+1} \partial_{x_{i+1}},\ \   i=1, \cdots, n-1.
 \nonumber
\eeqa

\noi
Furthermore, all the highest weight representations  of $su(n)$ can be 
obtained 
from (\ref{eq:sun}). Indeed, we can check using (\ref{eq:sun}) that
all the primitive vectors of the basic representations can be realized
in   terms of
antisymmetric products of the $x$'s (in this notation we have
$|e_i> = x_i$)

\beqa
\label{eq:rep_sun}
|\mu_{(1)}>&=& |e_1> =x_1, \nonumber \\
|\mu_{(2)}>&=&  |e_1 + e_2> =x_1 \wedge x_2, \nonumber \\
&\vdots& \\
|\mu_{(n-1)}>&=&|e_1 + \cdots+ e_{n-1}>= x_1 \wedge \cdots \wedge x_{n-1}. 
\nonumber
\eeqa

\noi
And correspondingly the representation $\D_\mu$ associated to the 
weight $\mu = \sum \limits_{i=1}^{n-1} p_i \mu_{(i)}$ with
$p_i \in \RR$ can be obtained from the highest weight
$|\mu> = \big(x_1\big)^{p_1} \cdots \big(x_1\wedge 
\cdots \wedge x_{n-1}\big)^{p_{n-1}}$
acting with the operators $E^{-\alpha_{(i)}}$ given in (\ref{eq:sun}).
When all the $p_i \in \NN$ 
we obtain a finite dimensional representation of $su(n)$ otherwise
the representation if infinite dimensional. Of course for $su(n)$ only the 
finite dimensional representations are unitary. Moreover, when $p_i \in \RR$
there is no guaranty that the representation can be exponentiated, namely that
they are representation of the Lie group $SU(n)$. Finally, notice that
the normalizations in (\ref{eq:sun}) and (\ref{eq:rep_sun}) are not 
the usual ones, but they are useful for to construct $F-$Lie algebras.
These last properties are also valid for all compact Lie algebras.

\subsection{$sp(2n)$}
Using, as for $su(n)$, the canonical basis of $\RR^n$, $e_i, i=1,\cdots, n$
the simple roots of $sp(2n)$  are

\beq
\label{eq:simple_sp2n}
\alpha_{(i)}=e_i-e_{i+1}, \ \ i=1,\cdots, n-1, 
\  \ \ \alpha_{(n)} = 2 e_n,
\eeq

\noi
$sp(2n)$ is not a simply-laced algebra (all roots do not have the same length).
The positive roots  are given  by 

\beq
\label{eq:roots_sp2n}
e_i \pm e_j, \ \ 1 \le i < j \le n, 
\ \ \ 2 e_i, \ \ 1 \le i \le n.
\eeq

Constructing the  {\bf 2n-}dimensional
representation associated to $\mu_{(1)}$ and
introducing $2n$ variables corresponding to the weight  of the 
representation $\D_{\mu_{(1)}}$
$x_i = |e_i>, x_{-i}=|-e_i>$ is is not difficult to define the operators
$E^{\pm \alpha_{(i)}}$ and $h_i$:

\beqa
\label{eq:sp2n}
E^{\alpha_{(i)}} &=& E^{e_i-e_{i+1}} = x_i \partial_{x_{i+1}}
+x_{-(i+1)} \partial_{x_{-i}},   \ \ 1 \le i \le n-1 \nonumber \\
E^{\alpha_{(n)}}&=& E^{2 e_n}= x_n \partial_{x_{-n}} \nonumber \\
E^{-\alpha_{(i)}} &=& E^{-e_i+e_{i+1}} = x_{i+1} \partial_{x_i}
+x_{-i} \partial_{x_{-(i+1)}},   \ \ 1 \le i \le n-1 \ \\
E^{-\alpha_{(n)}}&=& E^{-2 e_n}= x_{-n} \partial_{x_{n}} \nonumber \\
h_i&=& x_i \partial_{x_i}-x_{i+1} \partial_{x_{i+1}}
-x_{-i} \partial_{x_{-i}}+x_{-(i+1)} \partial_{x_{-(i+1)}},\ \   
i=1, \cdots, n-1.
 \nonumber \\
h_n&=&x_n \partial_{x_n}  -x_{-n} \partial_{x_{-n}} .\nonumber
\eeqa

\noi
The notations has been chosen in such a way that comparing the weights
of the variables $x > 0$ with the weights of the generators of
$sp(2n)$, the expression of the $E^\alpha$  can
 directly be read. This holds equally for
$su(n)$ and $so(n)$. 
Then, the  remaining generators can be calculated using the 
generators associated
to the primitives roots. Noticing that $E^{e_i-e_j}$ maps 
$|-e_i> \longrightarrow |-e_j>$ and $|e_j> \longrightarrow |e_i>$ we
get $E^{e_i-e_j} = a x_i \partial_{x_{j}} + b x_{-j} \partial_{x_{-i}}$.
The coefficients $a, b$ can be determined by multiple commutators. Indeed,
writing  ($j >i$~)
$j=i+k$, we get $e_i-e_{i+k}= \alpha_{(i)} + \cdots + \alpha_{(i+k-1)}$
and $ E^{e_i-e_j} = 
[E^{\alpha_{(i+k-1)}},\cdots[E^{\alpha_{(i+1)}},E{^{\alpha_{(i)}}}] \cdots ]$.
And similarly for the other generators. Now as for $su(n)$, all
representation can be obtained from (\ref{eq:sp2n}):

\beqa
\label{eq:rep_sp2n}
|\mu_{(1)}>&=& |e_1> = x_1, \nonumber \\
|\mu_{(2)}>&=& |e_1 + e_2> =x_1 \wedge x_2, \nonumber \\
&\vdots& \\
|\mu_{(n-1)}>&=& |e_1 + \cdots + e_{n-1}> = 
x_1 \wedge \cdots \wedge x_{n-1}, \nonumber \\
|\mu_{(n)}>&=&|e_1 + \cdots+ e_{n}> = 
x_1 \wedge \cdots \wedge x_{n}. \nonumber
\eeqa

\noi
To obtain all representations we proceed as for $su(n)$. Some remarks can be 
done at that point. For $sp(2n)$,  
$\left(\D_{\mu_{(1)}} \otimes\D_{\mu_{(1)}}\right)_{\mathrm{anti.~ sym}}
=\D_{\mu_{(2)}} \oplus \D_0$
($\D_0$ being the scalar representation) as can be seen directly from
(\ref{eq:rep_sp2n}) and (\ref{eq:sp2n}). Similar results hold for
$\D_{\mu_{(i)}}$.
 
\subsection{$so(2n+1)$}
This Lie algebra is the algebra dual of $sp(2n)$ and his 
simple roots are obtained
from the simple roots of $sp(2n)$ ( $(\alpha_{(i)})_{{so(2n+1)}} = 
2 (\alpha_{(i)})_{{sp(2n)}}/((\alpha_{(i)})_{{sp(2n)}})^2$). With the
same notations as for $su(n)$ and $sp(2n)$ we have the simple roots

\beq
\label{eq:simple_so2n+1}
\alpha_{(i)}=e_i-e_{i+1}, \ \ i=1,\cdots, n-1,
\  \ \ \alpha_{(n)} =  e_n.
\eeq

\noi
The positive roots  are given  by 

\beq
\label{eq:roots_so2n+1}
e_i \pm e_j, \ \ 1 \le i < j \le n, 
\ \ \  e_i, \ \ 1 \le i \le n.
\eeq

The vectorial representation (of dimension $2n+1$) allows to define
the set of variables $x_i = |e_i>, x_0 =|0>$ and $x_{-i} = |-e_i>$.
Constructing explicitly the vectorial representation  we obtain

\beqa
\label{eq:so2n+1}
E^{\alpha_{(i)}} &=& E^{e_i-e_{i+1}} = x_i \partial_{x_{i+1}}
+x_{-(i+1)} \partial_{x_{-i}},   \ \ 1 \le i \le n-1 \nonumber \\
E^{\alpha_{(n)}}&=& E^{e_n}= x_n \partial_{x_{0}}
+ x_{0} \partial_{x_{-n}} \nonumber \\
E^{-\alpha_{(i)}} &=& E^{-e_i+e_{i+1}} = x_{i+1} \partial_{x_i}
+x_{-i} \partial_{x_{-(i+1)}},   \ \ 1 \le i \le n-1 \ \\
E^{-\alpha_{(n)}}&=& E^{- e_n}= x_{0} \partial_{x_{n}}
+x_{-n} \partial_{x_0} \nonumber \\
h_i&=& x_i \partial_{x_i}-x_{i+1} \partial_{x_{i+1}}
-x_{-i} \partial_{x_{-i}}+x_{-(i+1)} \partial_{x_{-(i+1)}},\ \   
i=1, \cdots, n-1.
 \nonumber \\
h_n&=&x_n \partial_{x_n}  -x_{-n} \partial_{x_{-n}} .\nonumber
\eeqa

If consider the $p-$froms as for $su(n)$ and $sp(2n)$ we observe that

\beqa
\label{eq:rep_so2n+1}
|\mu_{(1)}>&=& |e_1> = x_1, \nonumber \\
|\mu_{(2)}>&=& |e_1 + e_2> =x_1 \wedge x_2, \nonumber \\
&\vdots& \\
|\mu_{(n-1)}>&=& |e_1 + \cdots+ e_{n-1}> = 
x_1 \wedge \cdots \wedge x_{n-1}, \nonumber \\
|2 \mu_{(n)}>&=&|e_1 + \cdots+ e_{n}> = 
x_1 \wedge \cdots \wedge x_{n}, \nonumber    
\eeqa

\noi
and all representations, except the spinorial one $D_{\mu_{(n)}}$ can
 be obtained
form (\ref{eq:rep_so2n+1}) and   (\ref{eq:so2n+1}). Then
if we want to obtain all representations
in terms of appropriate monomials another differential realization,
associated to $D_{\mu_{(n)}}$ has to be constructed
along the same lines
({\it constructing explicitly the spinorial representation}). 
However, if one consider
the highest weight states $|\mu_n> = (x_1 \wedge \cdots \wedge x_{n})^{1/2}$
one is able to define a primitive vector having all properties of the
primitive vector of the spinorial representation. But, in such a 
representation the operators $E^\alpha$ are not nilpotent. This means that
this representation is precisely a Verma module ${ \cal V}_{\mu_{(n)}}$. Then
${\cal V}_{\mu_{(n)}}$ has a unique maximal proper
sub-representation $M_{\mu_{(n)}}$ and the quotient 
${\cal V}_{\mu_{(n)}}/M_{\mu_{(n)}}$
is $\D_{\mu_{(n)}}$  (see {\it e.g. } \cite{vg,kr}) 
\footnote{If we use (\ref{eq:so2n+1}) for
$n=1$ {\it i.e. for $so(3)$} and we consider $(x_1)^{1/2}$ 
as a primitive vector of the spinorial representation  we can easily see
that for any $p > 0 ~~$ $\Big(E^{-\alpha_{(1)}}\Big)^p (x_1)^{1/2} \ne 0$ but
$E^{\alpha_{(1)}}\Big(E^{-\alpha_{(1)}}\Big)^2  (x_1)^{1/2} =0$. 
So the representation ${\cal V} = 
\left\{\Big(E^{-\alpha_{(1)}}\Big)^p (x_1)^{1/2}, p \ge 0 \right\}$
is precisely  a Verma module.
But it can be easily seen that 
$M= \left\{\Big(E^{-\alpha_{(1)}}\Big)^p (x_1)^{1/2}, p \ge 2 \right\}$
is the maximal sub-representation of ${\cal V}$ 
($\forall m \in M, E^{\alpha_{(1)}}(m),E^{-\alpha_{(1)}}(m), h_1(m) \in M $)
 and then $\D = {\cal V}/M$ is the two-dimensional spinorial representation.}.
But, constructing explicitly the representation ${\cal V}_{\mu_{(n)}}$ and
introducing at each step a new variable, $y_1,\cdots, y_{2^n}$, one is able
to construct the differential realization  of the spinorial representation.
Indeed, for any representation  $\D_{\mu}$ such a process can be equally
applied and the differential realization of $\D_{\mu}$ can be reached
straightforwardly.
Of course this is also possible for $su(n), sp(2n)$ and $so(2n)$.

\subsection{$so(2n)$}
The case of $so(2n)$ is similar to $so(2n+1)$ but in this case we have two
spinorial representations. With the same notations as before we introduce
the set of primitive roots

\beq
\label{eq:simple_so2n}
\alpha_{(i)}=e_i-e_{i+1}, \ \ i=1,\cdots, n-1,
\  \ \ \alpha_{(n)} =  e_{n-1} + e_n,
\eeq

\noi
and the positive roots  are given  by

\beq
\label{eq:roots_so2n}
e_i \pm e_j, \ \ 1 \le i < j \le n.
\eeq

From the {\bf 2n-}dimensional vector representation we obtain 
($x_i = |e_i>, x_{-i} = |-e_i>$)

\beqa
\label{eq:so2n}
E^{\alpha_{(i)}} &=& E^{e_i-e_{i+1}} = x_i \partial_{x_{i+1}}
+x_{-(i+1)} \partial_{x_{-i}},   \ \ 1 \le i \le n-1 \nonumber \\
E^{\alpha_{(n)}}&=& E^{e_{n-1}+ e_n}=  x_{n-1} \partial_{x_{-n}}
+x_{n} \partial_{x_{-(n-1)}}
 \nonumber \\
E^{-\alpha_{(i)}} &=& E^{-e_i+e_{i+1}} = x_{i+1} \partial_{x_i}
+x_{-i} \partial_{x_{-(i+1)}},   \ \ 1 \le i \le n-1  \\
E^{-\alpha_{(n)}}&=& E^{-e_{n-1} - e_n}= x_{-n} \partial_{x_{n-1}}
+x_{-(n-1)} \partial_{x_n} \nonumber \\
h_i&=& x_i \partial_{x_i}-x_{i+1} \partial_{x_{i+1}}
-x_{-i} \partial_{x_{-i}}+x_{-(i+1)} \partial_{x_{-(i+1)}},\ \
i=1, \cdots, n-1.
 \nonumber \\
h_n&=&x_{n-1} \partial_{x_{n-1}} +
x_n \partial_{x_n}  -x_{-n} \partial_{x_{-n}}
-x_{-(n+1)} \partial_{x_{-(n+1)}} .\nonumber
\eeqa

And, as for $so(2n+1)$ we observe that the $p-$forms are
 
\beqa
\label{eq:rep_so2n}
|\mu_{(1)}>&=& |e_1> = x_1, \nonumber \\
|\mu_{(2)}>&=& |e_1 + e_2> =x_1 \wedge x_2, \nonumber \\
&\vdots& \\
|\mu_{(n-1)} + \mu_{(n)}>&=& |e_1 + \cdots + e_{n-1}> =
x_1 \wedge \cdots \wedge x_{n-1}, \nonumber \\
|2 \mu_{(n)}>&=&|e_1 + \cdots + e_{n}> =
x_1 \wedge \cdots \wedge x_{n}. \nonumber
\eeqa

\noi
Then two more differential realizations (the two spinorial) allow
to construct all representations of $so(2n)$.
However, if we use (\ref{eq:so2n}) for all representations of $so(2n)$,
the spinorial representations will be  infinite dimensional Verma
modules.

To conclude this section recall once again that on the  one hand
the differential realization(s)
considered for $su(n), sp(2n), so(2n+1)$ and $so(2n)$ are just a convenience
and simply reproduce known results on Lie algebras. 
Of course similar expressions (in terms of bosonic or fermionic
oscillators) are given in standard text-books (see {\it e.g.} \cite{l})
but to my knowledge they
were not used to construct representations in a systematic
way from the fundamental one $\D_{\mu_{(1)}}$, except for $su(2), sl(2,\RR)$
and $sl(2,\CC)$ (see \cite{w, ggv-l}). Indeed, 
the advantage of the differential realization
(\ref{eq:sun}),(\ref{eq:sp2n}),(\ref{eq:so2n+1}) and (\ref{eq:so2n})
 is to permit constructing 
explicitly the representations in terms of appropriate 
monomials. For such representations, finite or infinite dimensional, 
 one, two or at maximum three realizations of the Lie algebra $g$ are just
needed.
In addition,  acting with (\ref{eq:sun}),
(\ref{eq:sp2n}), (\ref{eq:so2n+1}) or (\ref{eq:so2n}), on the states
of the corresponding representation $\D_\mu$ (of dimension $d$), and 
introducing new variables $y_1,\cdots, y_d$  at each step, permits the
construction of the differential realization of $\D_\mu$ directly form
$\D_{\mu_{(1)}}$. These differential realizations are also 
convenient to obtain representations neither bounded from below nor from
above (for instance, in the case of $su(n)$, the representation obtained with
$x_1^{p_1} \cdots x_n^{p_n}, p_i \not\in \NN$ has no primitive vector).
On the other hand, they are very useful to construct FSUSY
as shown in the next section.

\mysection{Fractional supersymmetry and Lie algebras}
For any Lie algebra $g$ and any representation $\D_\mu$,
one is able to construct an associated  FSUSY as it has been established
in \cite{vg}.  One possible solution is to consider the infinite dimensional
representation $\D_{\mu/F}$. At that point, the results of the previous 
section apply  and allow to define an $F-$Lie algebra associated to
$g, \D_\mu$ and $\D_{\mu/F}$. Indeed, this explicit procedure has only be
done for $so(1,2)$ \cite{fsusy3d,vg}. For higher rank Lie algebras
this construction was obtained abstractly in terms of Verma modules
\cite{vg}. But now, we observe that the differential realization
of $so(1,2)$ extends along the same lines for any Lie algebra. 
Then, it becomes possible to realize, as it has been done
for $so(1,2)$, an $F-$Lie algebra for any Lie algebra.
This explicit construction being analogous for any $g$ we just reproduce it
for the rank two Lie algebra $su(3)$ (weight diagrams and representations can
be represented graphically) but we have to keep in mind that this 
procedure, to construct an $F-$Lie algebra, 
is equally valid for {\it any Lie algebra} $g$.\\

Denote $\alpha, \beta$ the simple roots of $su(3)$ and $\gamma=\alpha+\beta$
the third positive root. Introduce, as in section 3, $x_1,x_2,x_3 >0$
of weight $|\mu_{(1)}> = x_1, |\mu_{(1)}-\alpha>=x_2, 
|\mu_{(1)}-\alpha -\beta>=x_3$. 
The generators takes the form

\beqa
\label{eq:su3}
&&\begin{array}{ll}
E^\alpha = x_1 \partial_{x_2}&E^{-\alpha} = x_2 \partial_{x_1} \cr
E^\beta = x_2 \partial_{x_3}&E^{-\beta} = x_3 \partial_{x_2} \cr
E^\gamma = x_1 \partial_{x_3}&E^{-\gamma} = x_3 \partial_{x_1} 
\end{array} \\
&& \ h_1=x_1 \partial_{x_1} - x_2 \partial_{x_2} \nonumber \\
&& \ h_2=x_2 \partial_{x_2}  - x_3 \partial_{x_3}. \nonumber
\eeqa

We would like to construct an $F-$Lie algebra associated to the three
dimensional representation $\D_{\mu_{(1)}}$
(this could have been done for any representation of  $su(3)$). 
In the  realization (\ref{eq:su3})
the vectorial representation writes

\beq
\label{eq:3}
\D_{\mu_{(1)}}= \left\{
\begin{array}{lll}
x_1&=|\mu_{(1)}>,& \cr
x_2&= |\mu_{(1)}-\alpha>&=E^{-\alpha}|\mu_{(1)}>, \cr 
x_3&=|\mu_{(1)}-\alpha -\beta>&=E^{-\beta}E^{-\alpha}|\mu_{(1)}>. 
\end{array}
\right.
\eeq  

So, we consider 

\beq
\label{eq:bos}
B=su(3) \oplus \D_{\mu_{(1)}}
\eeq

\noi
for the bosonic (graded zero part) of the $F-$Lie algebra. The natural
representation to define the ``$\mathrm{F}^{\mathrm{th}}-$root'' of 
$\D_{\mu_{(1)}}$ is  $\D_{\mu_{(1)}/F}$.
So, we take for the graded one part 

\beq
\label{eq:anyon}
A_1 = \D_{\frac{\mu_{(1)}}{F}}
\eeq

In the realization (\ref{eq:su3}) this representation writes

\beqa
\label{eq:3/F}
\D_{\frac{\mu_{(1)}}{F}} = \Big \{ \hskip -.7truecm
&&|\frac{\mu_{(1)}}{F} - n \alpha - p \beta> =
(x_1)^{1/F} \left(\frac{x_2}{x_1}\right)^n  \left(\frac{x_3}{x_2}\right)^p
\sim
\\
&& 
\left(E^{-\beta}\right)^p \left(E^{-\alpha}\right)^n 
|\frac{\mu_{(1)}}{F}>, 
n \in \NN, 0 \le p \le n  \Big\},
\nonumber
\eeqa

\noi
leading to the following weight diagram

\begin{figure}[!h]
 \epsfysize =7.cm
$$\epsffile{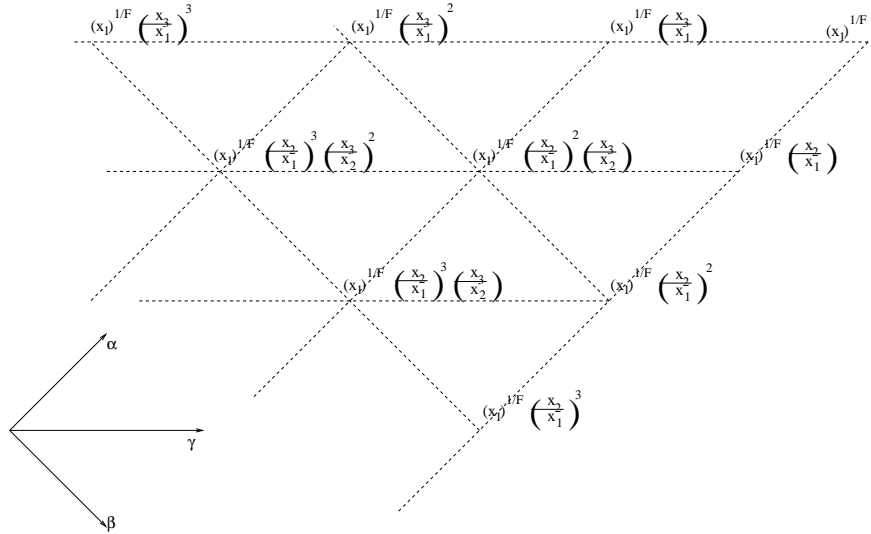}$$
\caption{\small{{Weight diagram for the $\D_{\mu_{(1)}/F}$ representation of
$su(3)$. The down-left corner represents the positive roots of $su(3)$.
The representations are infinite dimensional in the direction of $\alpha$
and $\gamma$, but finite dimensional in the direction $\beta$.}}}
\end{figure}

To define an $F-$Lie algebra associated to 
$su(3)\oplus\D_{\mu_{(1)}} \oplus \D_{\mu_{(1)}/F}$  we consider the
representation (reducible)

\beq
\label{eq:sf}
{\cal S}^F\left(\D_{\mu_{(1)}/F}\right)=
\left\{ \odot_{i=1}^F (x_1)^{1/F} \left(\frac{x_2}{x_1}\right)^{n_i}
\left(\frac{x_3}{x_2}\right)^{p_i}, \ \ n_i \in \NN, 0 \le p_i \le n_i
 \right\},
\eeq

\noi
with $\odot$ the symmetric tensorial product.

In we compare $x_1$, the primitive vector of $\D_{\mu_{(1)}}$ with 
$\otimes^F\left(x_1\right)^{1/F}$ we observe that these two vectors,
as primitive vectors, satisfy the same properties

\beq
\label{eq:primitive}
\begin{array}{ll}
h_1(x_1)=x_1,&h_1\Big(\otimes^F\left(x_1\right)^{1/F}\Big)=
\otimes^F\left(x_1\right)^{1/F}, \cr
h_2(x_1)=0,&h_2\Big(\otimes^F\left(x_1\right)^{1/F}\Big)=0, \cr
E^{\alpha,\beta,\gamma}(x_1)=0,&E^{\alpha,\beta,\gamma}
\Big(\otimes^F\left(x_1\right)^{1/F}\Big)=0.
\end{array}
\eeq

But now, if we construct the representation from these primitives vectors,
on the one hand we get $\D_{\mu_{(1)}}$ and on the other hand
the infinite dimensional representation

\beqa
\label{eq:verma}
\left<\otimes^F\left(x_1\right)^{1/F} \right> = \Big\{ \hskip - .7truecm
&&  |\mu_{(1)} - n \alpha - p \beta> = \\ 
&& \left(E^{-\beta}\right)^p \left(E^{-\alpha}\right)^n 
\Big(\otimes^F\left(x_1\right)^{1/F}\Big), 
n \in \NN, 0 \le p \le n  
\Big \}. \nonumber
\eeqa

\noi
But a direct calculation show that the following relations hold

\beqa
\label{eq:nil}
E^\alpha |\mu_{(1)}- 2 \alpha> =0 \nonumber \\
E^\alpha |\mu_{(1)}- 2 \alpha - \beta> =0 \\
E^\gamma |\mu_{(1)}- 2 \alpha - 2 \beta> =0.  \nonumber
\eeqa

\noi
This means that ${\cal V}_{\mu_{(1)}}=\left<\otimes^F\left(x_1\right)^{1/F} \right>$ is
a Verma module (the operator $E^\alpha$ is not nilpotent) and
$M=\left\{\left(E^{-\beta}\right)^p \left(E^{-\alpha}\right)^n 
\Big(\otimes^F\left(x_1\right)^{1/F}\Big), 
n \in \NN, 0 \le p \le n, (n,p) \ne (0,0), (1,0), (1,1) \right\}$ is the 
maximal proper sub-representation of ${\cal V}_{\mu_{(1)}}$
($\forall T \in su(3), \forall m \in M, T(m) \in M$). 
Then ${\cal V}_{\mu_{(1)}}/M$ and 
$\D_{\mu_{(1)}}$ are isomorphic.  Then, from ${\cal V}_{\mu_{(1)}} 
\subset
{\cal S}^F\left(\D_{\mu_{(1)}/F}\right)$ one can find an 
injection 
 $i: {\cal V}_{\mu_{(1)}} \longrightarrow {\cal S}^F\left(\D_{\mu_{(1)}/F}
\right)$.
Conversely, from $\D_{\mu_{(1)}} \equiv {\cal V}_{\mu_{(1)}}/M$  we can
define a surjection $\pi: {\cal V}_{\mu_{(1)}} \longrightarrow \D_{\mu_{(1)}}$. 
Consequently 
the following diagram 

$$\diagram{
{\cal S}^F\left(\D_{\mu_{(1)}/F}\right)
&\hfr{i}{}&{\cal V}_{\mu_{(1)}}
&\hfl{\pi}{}&\D_{\mu_{(1)}},
}$$

\noi
shows that we cannot define an $F-$Lie algebra in such a way
(because we cannot find a mapping
 from ${\cal S}^F\left(\D_{\mu_{(1)}/F}\right)$
into $\D_{\mu_{(1)}}$ as stated in property 2 of $F-$Lie algebras, see
section 2).
Indeed, in \cite{fsusy3d} FSUSY in $1+2$ dimensions was constructed along these
lines but we did not have the structure of $F-$Lie algebra.

To obtain an $F-$Lie algebra some constraints have to be introduced.
Following \cite{vg}, we define ${\cal F}$ the vector space of functions
on  $x_1,x_2,x_3 >0$. The multiplication map $m_n: {\cal F} \times 
\cdots \times {\cal F} \longrightarrow {\cal F}$ given by
$m_n(f_1,\cdots,f_n) = f_1 \cdots f_n$ is multilinear and totally symmetric.
Hence, it induces a map $\mu_F$ from ${\cal S}^F\left({\cal F} \right)$
into ${\cal F}$. Restricting to ${\cal S}^F\left({\D_{\mu_{(1)}}} \right)$ one
gets

\beqa
\label{eq:sfred}
\mu_F:{\cal S}^F\left({\D_{\mu_{(1)}}} \right) &\longrightarrow&
{\cal S}_{{\mathrm {red}}}^F\left({\D_{\mu_{(1)}}} \right) \nonumber \\
\odot_{i=1}^F (x_1)^{1/F} \left(\frac{x_2}{x_1}\right)^{n_i}
\left(\frac{x_3}{x_2}\right)^{p_i} &\longrightarrow&
x_1 \left(\frac{x_2}{x_1}\right)^{\sum \limits_{i=1}^F n_i}
\left(\frac{x_3}{x_2}\right)^{\sum \limits_{i=1}^F p_i}.
\eeqa

We observe that 
${\cal S}_{{\mathrm {red}}}^F\left({\D_{\mu_{(1)}}} \right)
=\left\{
x_1 \left(\frac{x_2}{x_1}\right)^n  \left(\frac{x_3}{x_2}\right)^p, 
n \in \NN, 0 \le p \le n  \right\}
 \supset
\D_{\mu_{(1)}}$, meaning that we can find an injection $i^\prime$ from
$\D_{\mu_{(1)}}$ into ${\cal S}_{{\mathrm {red}}}^F\left({\D_{\mu_{(1)}}}
\right)$.
This representation is reducible but indecomposable.
Namely, we cannot
find a complement of $\D_{\mu_{(1)}}$ in 
${\cal S}_{{\mathrm {red}}}^F\left({\D_{\mu_{(1)}}} \right)$ stable under
$su(3)$. For instance $x_1 \left(\frac{x_2}{x_1}\right)^2$ is such
that $E^\alpha \left(x_1 \left(\frac{x_2}{x_1}\right)^2\right)= 2 x_2$,
but $E^{-\alpha}(x_2)=0$.
As before we observe that the diagram

$$\diagram{
{\cal D}_{\mu_{(1)}}
 &\hfl{i^\prime}{}&  {\cal S}^F\left( {\cal D}_{\mu_{(1)}/ F}\right)_
{\mathrm {red}}
&\hfr{\mu_F}{}&{\cal S}^F\left( {\cal D}_{\mu_{(1)} / F}\right) 
}$$

\noi
leads to the same conclusion on the structure of $F-$Lie algebra.
With these simple observations we can conclude as in \cite{vg} that
we cannot define an $F-$Lie algebra with $B=su(3) \oplus \D_{\mu_{(1)}}$.
To obtain such a structure, one possible solution is
to extend $\D_{\mu_{(1)}}$ into an infinite dimensional representation. 
For instance,
 
\beq
\label{eq:flie}
\left(su(3) \oplus 
{\cal S}^F\left( {\cal D}_{\mu_{(1)}/ F}\right)_{\mathrm {red}}\right)
\oplus {\cal D}_{\mu_{(1)}/ F}
\eeq

\noi
has a structure of $F-$Lie algebra
(a similar structure could have been defined with
$\Big(su(3) $ $ \oplus {\cal V}_{\mu_{(1)}} \Big) \oplus 
{\cal D}_{\mu_{(1)}/ F}$).

The problem to construct an $F-$Lie algebra associated to $su(3),
\D_{\mu_{(1)}}, \D_{\mu_{(1)}/F}$ is basically related to the fact that
we  would like to relate a finite dimensional representation
$\D_{\mu_{(1)}}$ with an infinite dimensional one
$\D_{\mu_{(1)}/F}$.  One possible solution, as we just have seen, is
to extend $\D_{\mu_{(1)}}$ into a infinite dimensional (reducible but 
indecomposable) representation 
${\cal S}^F\left( {\cal D}_{\mu_{(1)}/ F}\right)_{\mathrm {red}}$.
This procedure being quite general, works similarly for any Lie algebra
$g$.

\mysection{Conclusion}
Under some assumptions symmetries beyond supersymmetry can be constructed.
Fractional supersymmetry and $F-$Lie algebras are a  possible solutions. 
In this note,  we have shown  that differential realization(s) of Lie
algebras is(are) an useful tool(s) to construct explicitly a structure of
$F-$Lie algebra associated to any representation $\D_\mu$ of any Lie 
algebra $g$. The basic point is to consider the infinite dimensional
representation $\D_{\mu/F}$. We have shown, that considering an
infinite dimensional (reducible but indecomposable) representation
extending $\D_{\mu}$ enables us to construct an $F-$Lie algebra
(this solves the problem related to the fact that, in general,
$\D_\mu$ is finite dimensional, although $\D_{\mu/F}$ is infinite
dimensional). Another possible solution, if to extend the Lie algebra
$g$ into a infinite dimensional Lie algebra. When $g=so(1,2)$ this
algebra reduces to the centerless Virasoro algebra \cite{vg}.
Another advantage of the differential realization of section 3 is the
possibility to construct explicitly, in a differential way, this
infinite dimensional algebra (this will be done elsewhere, 
at least for $su(n)$).\\
Finally, we would like to conclude that for $g=so(1,2)$ unitary 
representations have been constructed \cite{fsusy3d}. It has been
observed that it is a symmetry which acts on relativistic anyons
\cite{an}.
The question of  the interpretation of FSUSY in higher dimensional 
space-time is still open.

\vskip1truecm
{\bf Acknowledgements:} J'exprime mes plus vifs remerciements
aux organisateurs  du R\'eseau de Physique Th\'eorique du Maroc, et en
particulier \`a A. El Hassouni et E. H. Saidi, pour leur invitation \`a
participer  au workshop
{\it Non Commutative Geometry and Superstring Theory},
16 et 17 Juin 2000, Rabat.
 

\end{document}